\documentclass[
reprint,
superscriptaddress,
amsmath,amssymb,
aps,
pra,
showkeys
]{revtex4-2}

\usepackage{graphicx}
\usepackage{dcolumn}
\usepackage{bm}
\usepackage{float}
\usepackage{upgreek}
\usepackage[T1]{fontenc}
\usepackage[utf8]{inputenc}
\usepackage{babel}
\usepackage{csquotes}

\begin{document}
\preprint{APS/123-QED}

\title{Temperature dependence of electron viscosity\\ in superballistic GaAs point contacts}

\author{Daniil I. Sarypov}
\email{d.sarypov@g.nsu.ru}
\author{Dmitriy A. Pokhabov}
\author{Arthur G. Pogosov}
\author{Evgeny Yu. Zhdanov}
\affiliation{Rzhanov Institute of Semiconductor Physics SB RAS, Novosibirsk, 630090, Russia}
\affiliation{Novosibirsk State University, Novosibirsk, 630090, Russia}
\author{Andrey A. Shevyrin}
\affiliation{Rzhanov Institute of Semiconductor Physics SB RAS, Novosibirsk, 630090, Russia}
\author{Askhat K. Bakarov}
\author{Alexander A. Shklyaev}
\affiliation{Rzhanov Institute of Semiconductor Physics SB RAS, Novosibirsk, 630090, Russia}
\affiliation{Novosibirsk State University, Novosibirsk, 630090, Russia}

\date{\today}

\begin{abstract}
Electron transport in suspended and non-suspended GaAs point contacts (PCs) of different widths is experimentally studied. The superballistic contribution to the conductance, that demonstrates a distinctive quadratic dependence on the PC width and temperature growth, is extracted from the experiment. The studied PCs are shown to be described in the framework of hydrodynamic electron flow a in wide temperature range. At low temperatures, $T$, the viscosity is found out to obey the law $1/T^2$ expected for 2D systems, while at higher temperatures it has the dependence $1/T$. Similar measurements performed after the suspension of PCs, i.e. their separation from substrate, show that the electron viscosity reduces in the whole temperature range, that indicates an enhanced electron-electron interaction in suspended structures.

\end{abstract}

\maketitle

\section{Introduction}

Electron-electron (e-e) interaction has a significant impact on the electron transport in mesoscopic semiconductor structures and, in some cases, its role becomes predominant \cite{meyer2008, ho2018, thomas2000, debray2009, pokhabov2018, pokhabov2021}.  At the same time, effects associated with e-e interaction, because of their many-body origin, are difficult to describe. Theoretical models usually use approximations, whose applicability is not always obvious, or involve numerical calculations that lack the desired generality and predictability. As a result, direct theoretical calculations do not provide a reliable physical concept which could promptly explain, predict and engineer the e-e interaction effects in electron transport. Such concept is given by an intuitive analogy of the collective electron motion with the hydrodynamic flow of a viscous liquid \cite{gurzhi1968, andreev2011, narozhny2015, alekseev2016, principi2016, lucas2018, narozhny2019, alekseev2020, polini2020, narozhny2022}, which is applicable in cases when the momentum conserving e-e scattering dominates over momentum relaxing scattering mechanisms. Although the idea of hydrodynamic approach, as well as the prediction of a non-obvious viscosity-related effect of the resistance drop with increasing temperature, have been developed for quite a long time \cite{gurzhi1962}, the first experimental observations of hydrodynamic effects appeared more recently \cite{dejong1995}.

Hydrodynamic effects are experimentally observable in sufficiently clean mesoscopic systems with a large momentum relaxing length $l$ (including electron-phonon and electron-impurity scattering) at temperatures that are low enough so that the influence of electron-phonon scattering on the electron transport is negligible, but high enough for the e-e scattering length $l_\mathrm{ee}$ that become the smallest lengthscale in the system. An example of the system perfectly suitable for the electron hydrodynamic research is two-dimensional electron gas (2DEG) in graphene, where $l$ can reach tens of microns at low $T$. Recent studies in the graphene-based Hall bars in vicinity geometry report the observation of unusual negative local resistance related to the electron viscosity \cite{bandurin2016, torre2015, levitov2016}. Another clean enough system appropriate for the electron hydrodynamics studies is a 2DEG in GaAs. Nowadays, there are many experimental contributions devoted to the observation of hydrodynamic effects in GaAs \cite{gusev2018, gusev2020, raichev2020, keser2021, gupta2021, ginzburg2023}.

For a complete hydrodynamic description of the electron transport it is essential to study in detail the relation between the hydrodynamic quantities and the parameters characterizing transport phenomena. As an example, the electron viscosity $\nu$ is traditionally considered proportional to the e-e scattering timescale $\tau_\mathrm{ee}=l_\mathrm{ee}/v_\mathrm{F}$:
\begin{equation}
    \label{eq:nu_tau}
    \nu=\frac{1}{4}v^2_\mathrm{F}\tau_\mathrm{ee},
\end{equation}
where $v_\mathrm{F}$ is the Fermi velocity \cite{gurzhi1968}. The timescale $\tau_\mathrm{ee}$ was the subject of theoretical investigations \cite{chaplik1971, giuliani1982} that showed that, for 2D electron systems, 
\begin{equation}
    \label{eq:tau}
    \tau_\mathrm{ee}\propto\frac{1}{T^2\ln{\left(\frac{E_\mathrm{F}}{k_\mathrm{B}T}\right)}}.
\end{equation}
However, both of these formulae \eqref{eq:nu_tau} and \eqref{eq:tau} are applicable only at very low temperatures relative to the Fermi energy ($k_\mathrm{B}T\ll E_\mathrm{F}$), while Eq.~\eqref{eq:nu_tau} can be generalized to higher temperatures (see, e.g., Ref.~\cite{jensen1980}):
\begin{equation}
    \label{eq:nu_general}
   \nu=\frac{v^2_\mathrm{F}}{4E^2_\mathrm{F}}\int\limits_0^{+\infty}{\tau_\mathrm{ee}(\varepsilon)\left(-\frac{\partial f_0}{\partial \varepsilon}\right)\varepsilon^2\,d\varepsilon}, 
\end{equation}
where $f_0$ is an equilibrium distribution function of electrons. 
\begin{figure*}[ht!]
    \centering
    \includegraphics{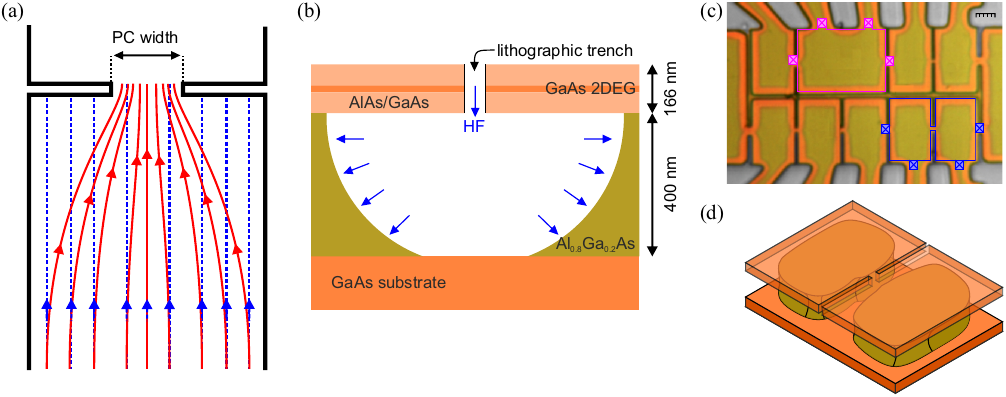}
    \caption{Schematic representation of: (a) electron streamlines through PC in ballistic (dashed lines) and hydrodynamic (solid lines) regimes and (b) the GaAs/AlGaAs heterostructure with the sacrificial Al$_{0.8}$Ga$_{0.2}$As layer which can be removed by a wet etching through lithographic trenches. (c) Optical micrograph of one of the experimental samples. Scale bar is 5~$\mu$m. Measurements schemes of the reference region and one of the PCs are superimposed on the top of image. (d) Schematic image of a suspended PC.}
    \label{fig:first}
\end{figure*}
Note that \eqref{eq:nu_general} coincides with  \eqref{eq:nu_tau} at $k_\mathrm{B}T\ll E_\mathrm{F}$, but, at $k_\mathrm{B}T\sim E_\mathrm{F}$, the quantities  $\nu$ and $\tau_\mathrm{ee}$, in general, differently depend on $T$. In graphene-based systems, due to large Fermi energy $E_\mathrm{F}\gtrsim 100$~meV, relations \eqref{eq:nu_tau} and \eqref{eq:tau} are applicable up to high enough temperatures ($T\lesssim100$~K). Therefore, the theoretically predicted dependence \eqref{eq:tau} finds its confirmation in various experiments, such as transverse magnetic focusing of ballistic electrons \cite{lee2016, berdyugin2020} and superballistic viscous flow of electrons through PC \cite{krishnakumar2017}. This is one of the reasons why the idea that $\nu\propto\tau_\mathrm{ee}\propto1/T^2$ has become a paradigm. However, in case of GaAs 2DEG with a typical concentration $n\sim10^{11}$~cm$^{-2}$ and $E_\mathrm{F}\sim10$~meV, the dependence \eqref{eq:tau} can be broken already at $T\sim10$~K, as noted in Ref.~\cite{keser2021}. Furthermore, the temperature dependence of the correction to the PC conductance due to the e-e interaction measured in GaAs 2DEG at $E_\mathrm{F}=3$~meV shows a divergence with the generally accepted dependence \eqref{eq:tau} \cite{melnikov2012}. Thus, up to date, the question of the temperature dependence of the electron viscosity in thr GaAs 2DEG remains open. This work is aimed at solving this problem experimentally.

Here we present the electron viscosity measurement in the GaAs 2DEG in the wide temperature range from $k_\mathrm{B}T=0.014E_\mathrm{F}$ to $k_\mathrm{B}T=0.3E_\mathrm{F}$. Besides, the electron viscosity is controlled not only by changing the temperature, but also by suspending the samples, i.e. separating them from the substrate. We show that suspending the samples leads to the reduction of $\nu$, caused by the e-e interaction enhancement in suspended structures, due to the removal of a part of the screening medium with a high dielectric constant from under the structure \cite{pogosov2022, chaplik1972}. At high temperatures, the electron viscosity $\nu(T)$, as a function of temperature, is found out to show the dependence that distinguishes it from Eq.~\eqref{eq:tau}, as expected for 2D systems.

The viscosity measurement method is based on the effect of the resistance reduction with increasing temperature, which is similar to the Gurzhi effect \cite{gurzhi1962}. The resistance decrease in our case is caused by the superballistic conductance effect, predicted theoretically \cite{guo2017} and also observed experimentally in graphene \cite{krishnakumar2017}. This effect can be illustrated by the difference in the electron current streamlines flowing through the PC in ballistic and hydrodynamic regimes. In the ballistic regime the current streamlines are straight, directed from source to drain (see Fig.~\ref{fig:first}~(a)) and the PC conductance $G_\mathrm{pc}$ is determined by the number of modes transmitted through it, while each one contributes as $2e^2/h$ \cite{sharvin1965}:
\begin{equation}
    \label{eq:Gball}
    G_\mathrm{pc}=G_\mathrm{ball}=\frac{2e^2}{h}\frac{k_\mathrm{F}w}{\pi},
\end{equation}
where $k_\mathrm{F}$, $w$ are the Fermi quasi-wavevector and the PC width, correspondingly. The case of the hydrodynamic regime is qualitatively different. In this regime, the current streamlines that are initially directed at the PC walls bend and pass through it, thus increasing the current and, consequently, the conductance $G_\mathrm{pc}$. Guo \textit{et al.} showed \cite{guo2017} that the described conductance increase caused by the e-e scattering takes an additive form:
\begin{equation}
   \label{eq:Gpc} G_\mathrm{pc}=G_\mathrm{ball}+G_\mathrm{vis},
\end{equation}
where viscous contribution $G_\mathrm{vis}$ is inversely proportional to electron viscosity $\nu$:
\begin{equation}
    \label{eq:Gvis}
    G_\mathrm{vis}=\frac{\pi}{32}\frac{e^2}{h}\frac{k_\mathrm{F}v_\mathrm{F}w^2}{\nu}.
\end{equation}
It is worth noting that the similar correction to the conductance $\Delta G\propto w^2 T$ due to the e-e scattering is predicted in Ref.~\cite{nagaev2008} without involving the hydrodynamic formalism.

The relation between $G_\mathrm{vis}$ and $\nu$ in Eq.~\eqref{eq:Gvis} makes it possible to extract the temperature dependence $\nu(T)$ from the measurements of $G_\mathrm{vis}(T)$. Such measurements were performed early in graphene-based PC \cite{krishnakumar2017} up to the temperatures $k_\mathrm{B}T\approx 0.16E_\mathrm{F}$, where the dependence $\nu\propto 1/T^{2}$ is observed in an almost entire temperature range, that is in agreement with the theoretical predictions at $k_\mathrm{B}T\ll E_\mathrm{F}$. Nevertheless, even in the mentioned study, there are deviations from the dependence $1/T^2$ at high temperatures $T\gtrsim200$~K. Similar measurements performed in the GaAs PC \cite{ginzburg2021} contains only an estimation of $l_\mathrm{ee}=v_\mathrm{F}\tau_\mathrm{ee}$ and has no evidence of the quadratic dependence $G_\mathrm{vis}(w)$ predicted by Eq.~\eqref{eq:Gvis}. In our work, in addition to the measurements of the electron viscosity temperature dependence, we analyze the viscous contribution $G_\mathrm{vis}$ dependence on the PC width $w$ and check its quadratic scaling, as predicted by Eq.~\eqref{eq:Gvis}.

\section{Materials and Methods}
\label{section:mmethods}

Experimental samples are created from the GaAs/AlGaAs heterostructure with a 2DEG grown by the molecular-beam epitaxy on a GaAs substrate (Fig.~\ref{fig:first}~(b)). The heterostructure is a short-period AlAs/GaAs superlattice 166~nm thick with a 13~nm GaAs layer forming a symmetric quantum well for electrons. Doping the heterostructure with the Si donor $\updelta$-layers located symmetrically on both sides of the potential well makes it possible to fill the quantum well with electrons forming the 2DEG. The heterostructure also contains the sacrificial 400~nm thick Al$_{0.8}$Ga$_{0.2}$As layer which can be removed by a selective wet etching in order to separate the studied structures from the substrate, i.e. to suspend them. The mobility and concentration of electrons at $T=4$~K are $\mu=2\times10^{6}$~cm$^{2}$/$(\rm{V}\cdot\rm{s})$ and $n=7\times10^{11}$~cm$^{-2}$, respectively. The momentum relaxing length at these parameters is $l=25$~$\upmu$m, and the Fermi energy is $E_\mathrm{F}=25$~meV. The heterostructure also contains low-mobility X-valley electrons, with a characteristic concentration of $10^{10}$~cm$^{-2}$, localized on Si donors; these electrons do not contribute to the conductance at low temperatures, but smooth fluctuations of the electrostatic potential of impurities, thus, increasing the 2DEG mobility. Such an application of X-valley electrons was proposed in Ref.~\cite{friedland1996}.

Several identical Hall bars with dimensions $50\times30$~$\upmu\text{m}^2$ were created by means of photolithography and a subsequent reactive ion etching. Then electron lithography was used to form a series of PCs of different lithographic width $w_\mathrm{lith}$ varying from 0.7 to 1.5~$\upmu$m with the step of 0.2~$\upmu$m  (Fig.~\ref{fig:first}~(c)). The effective PC width $w=w_\mathrm{lith}-w_\mathrm{depl}$, with the depletion layer thickness $w_\mathrm{depl}$ considered, is lower than the lithographic width. Effective PC widths are determined in Sec.~\ref{section:results} from low-$T$ resistance measurements. Note that all the created samples demonstrate the reproducibility of the obtained results.

The experimental samples also include a reference region, i.e. a region without the PC (see Fig.~\ref{fig:first}~(c)), which reproduces the geometry of the regions adjacent to the PC. As is shown below in Sec.~\ref{section:results}, the phonon and impurity contributions to the measured resistance can be determined by measuring the temperature dependence of the reference region resistance and thus effects directly related to the electron viscosity can be extracted from the experiment.

In order to suspend the studied PCs, the sacrificial Al$_{0.8}$Ga$_{0.2}$As layer was selectively etched by a hydrofluoric acid solution. The solution was introduced through the 200~nm deep lithographic trenches. Such depth is essential for the etchant to reach the sacrificial layer. For the detailed description of the suspension technology see, e.g., Ref.~\cite{pogosov2022}. The suspended area size is limited to the order of 4~$\upmu$m to avoid the sagging of the suspended structure and its undesirable contact with the substrate. The suspended PC obtained as a result of the suspension is schematically shown in Fig.~\ref{fig:first}~(d). The suspension leaves the 2DEG mobility and concentration almost unchanged, and that is confirmed by the Hall measurements.

The experimental samples were equipped with the Au/Ni/Ge ohmic contacts. The PC resistance was measured according to the standard four-terminal scheme in the cryostate in the temperature range from 4 to 90~K by means of the lock-in technique at the frequency of 70~Hz and the excitation current amplitude of 100~nA. The measurements were first performed before the suspension and then repeated after the suspension.

\begin{figure}[ht!]
    \centering
    \includegraphics{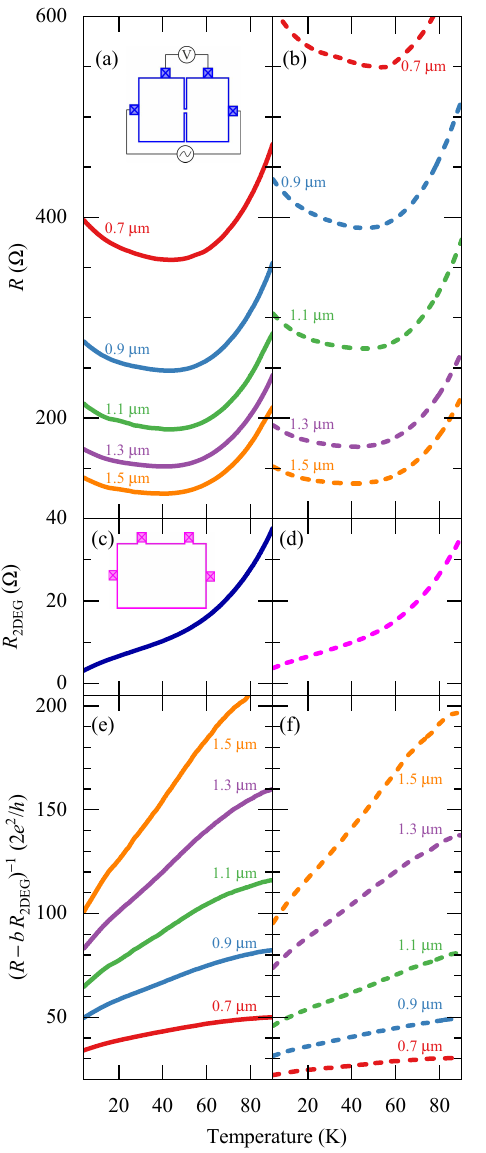}
    \caption{Temperature dependence of PC resistance $R$ including the resistance of adjacent regions (a-b), reference region resistance $R_{\mathrm{2DEG}}$ (c-d) and the PC conductance $\left(R-b R_{\mathrm{2DEG}}\right)^{-1}$ (e-f). Solid and dashed curves on all panels correspond to a measurement before and after the suspension, correspondingly. The numbers on the graphs denote a lithographic width of a PC. The measurement schemes are shown in the insets.}
    \label{fig:resistance}
\end{figure}
\section{Results and Discussion} \label{section:results}
In Fig.~\ref{fig:resistance}~(a-b) is the temperature dependence of the PCs resistance $R(T)$ measured before (solid curves) and after (dashed curves) the suspension according to the scheme shown in the inset to Fig.~\ref{fig:resistance}~(a). It is clearly seen that the resistance first decreases at temperatures $\lesssim 40$~K for all PCs and then monotonically increases. The observed resistance drop with increasing temperature is similar to the Gurzhi effect \cite{gurzhi1962}.

The measured quantity $R$ consists of the PC resistance and the resistance of the regions adjacent to it $R_\mathrm{c}$, associated with the scattering on phonons and impurities:
\begin{equation}
    R=R_\mathrm{c}+G_\mathrm{pc}^{-1},
    \label{eq:measured}
\end{equation}
where $G_\mathrm{pc}$ is the PC conductance determined by Eq.~\eqref{eq:Gpc}. Thus, according to Eq.~\eqref{eq:Gpc}, the e-e interaction leads to the drop of $R$ while the electron-phonon interaction causes its growth with increasing temperature. To extract the viscous contribution $G_{\mathrm{vis}}$ from the experiment, the adjacent regions resistance $R_\mathrm{c}$ was subtracted from $R$. Since all the electron-phonon and electron-impurity collisions emerge in the PC adjacent regions, the resistance $R_\mathrm{c}$ and the reference region resistance $R_\mathrm{2DEG}$ have the same temperature dependence. (Sec.~\ref{section:mmethods}). The dependence $R_\mathrm{2DEG}(T)$ shown in Fig.~\ref{fig:resistance}~(c-d) demonstrates the monotonic character typical of metals. We suppose that $R_\mathrm{c}$ is related to $R_\mathrm{2DEG}$ as $R_\mathrm{c}=b R_\mathrm{2DEG}$, where the geometric factor $b$ takes into account the PC boundaries.

A possible method of determining the factor $b$ is a numeric computation of $R_\mathrm{c}$ by the self-consistent solution of the Laplace equation and the continuity equation in the particular PC geometry with the corresponding boundary conditions, as it was performed in Ref.~\cite{krishnakumar2017}. This calculation contains a phenomenological parameter, namely, the scattering time $\tau$ that accounts all the momentum non-conserving collisions including scattering at the sample boundaries. The parameter $\tau$ can not be calculated accurately, and that introduces a significant error to the subsequent analysis. We propose the factor $b$ determination method that is based only on the experimental data without any fitting parameters. The first term $R_\mathrm{c}(T)$ in Eq.~\eqref{eq:measured}, as a function of $T$, grows unlimitedly with increasing $T$ that caused is by the increase of the electron-phonon scattering rate, while the second term $1/G_\mathrm{pc}(T)$, according to the Eq.~\eqref{eq:Gpc}, decreases being limited from below ($G_\mathrm{pc}>0$). It means that the $T$-derivative of $1/G_\mathrm{pc}$ tends to zero at high enough temperatures. So the observed growth of $R$ (Fig.~\ref{fig:resistance}-(a-b)) is mainly due to the growth of $R_\mathrm{c}(T)=bR_\mathrm{2DEG}(T)$. Therefore, differentiating Eq.~\eqref{eq:measured} by $T$, we obtain:
$$\frac{dR}{dT}=b\frac{dR_\mathrm{2DEG}}{dT}+\frac{d}{dT}G_\mathrm{pc}^{-1}\approx b\frac{dR_\mathrm{2DEG}}{dT}.$$
Thus, the geometric factor $b$ can be obtained directly from the experimental data as follows:
\begin{equation}
    b \approx \left. {\left( {{{\frac{{dR}}{{dT}}} \mathord{\left/
 {\vphantom {{\frac{{dR}}{{dT}}} {\frac{{dR_{{\rm 2DEG}} }}{{dT}}}}} \right.
 \kern-\nulldelimiterspace} {\frac{{dR_{{\rm 2DEG}} }}{{dT}}}}} \right)} \right|_{T = 90\;{\rm K}} .
  \label{eq:factor}
\end{equation}
The factor $b$ determined this way takes the values from 4.4 to 7.2 for different PCs. Note that the numeric computation of $b$ similar to that in Ref.~\cite{krishnakumar2017} with $\tau=m\mu/e$ ($m$ is the effective electron mass in GaAs) gives the factor $b$ values underestimated 1.2-1.7 times. In a further analysis we will use the $b$ values obtained from Eq.~\eqref{eq:factor}. 

In Fig.~\ref{fig:resistance}~(e-f) is the PC conductance without the adjacent regions contribution $\left(R-bR_\mathrm{2DEG}\right)^{-1}=G_\mathrm{pc}$. It is seen that, after the subtraction of the adjacent regions resistance associated with the scattering on phonons and impurities, the conductance of all PCs monotonically grows in the whole temperature range. Such behavior shows that e-e collisions lead to the conductance growth with increasing temperature and confirms the presence of the superballistic conductance (Eq.~\eqref{eq:Gpc}). Indeed, the ballistic conductance $G_\mathrm{ball}$ is $T$-independent while, according to Eq.~\eqref{eq:Gvis}, the viscous contribution $G_\mathrm{vis}(T)\propto \nu^{-1}$ monotonically grows with temperature. Note also that the temperature dependence observed in Fig.~\ref{fig:resistance}~(e-f) looks other than parabolic, that can be expected since $\tau_\mathrm{ee}$, determining the viscosity according to the conventional relation \eqref{eq:nu_tau}, has the dependence $1/\tau_\mathrm{ee}\propto T^{2}\ln{T}$ \cite{chaplik1971, giuliani1982}. Instead, we see an almost linear dependence in the wide temperature range, except for the low-$T$ and high-$T$ limits, where a deviation from the linear dependence becomes noticeable. A detailed analysis of the temperature dependence will be provided below.

Since the viscous term $G_\mathrm{vis}$ is much less than the ballistic term $G_\mathrm{ball}$ at low temperature, the PC conductance is determined by $G_\mathrm{ball}$ at $T=4$~K. It allows us to obtain effective widths $w$ from Eq.~\eqref{eq:Gball} and, consequently, the ballistic contribution to the conductance $G_\mathrm{ball}$ for each PC. The depletion layer width $w_\mathrm{depl}$ slightly varies in the range 0.06-0.1~$\mu$m before the suspension and 0.3-0.4~$\mu$m after the suspension, where larger $w_\mathrm{depl}$ values correspond to a narrower PC. Differences in $w_\mathrm{depl}$ among various PCs may be caused by a different screening efficiency due to the lower electron concentration inside the narrow PC, compared with the wider one.

\begin{figure}[ht!]
    \centering
    \includegraphics{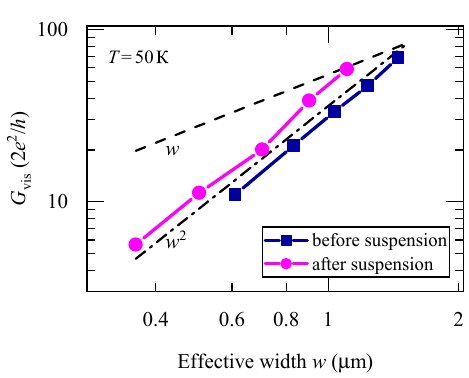}
    \caption{Viscous contribution $G_\mathrm{vis}$ to the PC conductance as a function of effective PC width $w$ for $T=50$~K on the log-log scale before (shown by squares) and after (shown by circles) the suspension. Dashed and dash-dotted lines show $w$ and $w^2$ dependence, correspondingly.}
    \label{fig:Gvis_w}
\end{figure}

Further, we subtracted the conductance $G_\mathrm{ball}$ from $G_\mathrm{pc}$ to obtain the viscous contribution $G_\mathrm{vis}=G_\mathrm{pc}-G_\mathrm{ball}$. The latter is shown in Fig.~\ref{fig:Gvis_w} 
as a function of the effective PC width $w$ at $T=50$~K. At this temperature the values of the ballistic and viscous contributions to the PC conductance become comparable. Note that $G_\mathrm{vis}$ is proportional to $w^2$ in contrast to the $G_\mathrm{ball}$ which is proportional to $w$, that is in agreement with Eq.~\eqref{eq:Gvis} and similar measurements in graphene-based PCs \cite{krishnakumar2017}. Thus, with the correctly determined parameters $b$ and $w_\mathrm{depl}$, we extracted the bare viscous contribution to the PC conductance from the experimental data.

\begin{figure}[h!]
    \centering
    \includegraphics{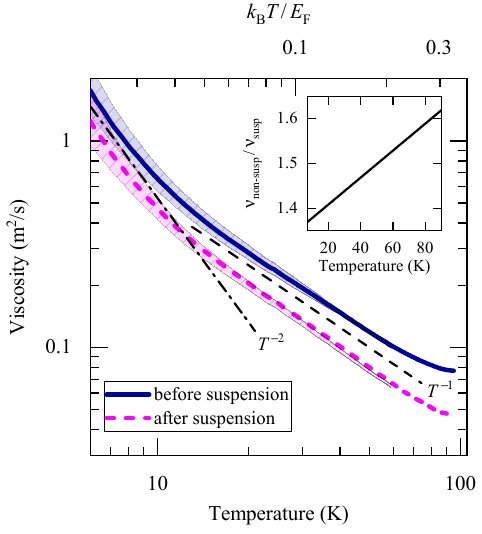}
    \caption{Temperature dependence of electron viscosity on the log-log scale before (solid curve) and after (dashed curve) the suspension. Filled regions show a possible 1\% inaccuracy in the ballistic conductance measurement. Dashed and dash-dotted lines show $1/T$ and $1/T^2$ dependence, correspondingly. The ratio of viscosities before and after the suspension as a function of temperature is shown in the inset.}
    \label{fig:viscosity}
\end{figure}
Finally, using the measured temperature dependence of the viscous conductance $G_\mathrm{vis}(T)$ and Eq.~\eqref{eq:Gvis} we get the temperature dependence of viscosity $\nu(T)$ shown in Fig.~\ref{fig:viscosity}. Note that, at $T\lesssim 15$~K ($k_\mathrm{B}T/E_\mathrm{F} \lesssim 0.05$), the viscosity falls as $\nu \propto 1/T^{2}$, while it obeys the law $\nu \propto 1/T$ at higher temperatures. Such deviation from the commonly accepted dependence $\nu\propto\tau_\mathrm{ee}\propto1/T^{2}$ can be explained by the fact that, in such conditions, the temperature smearing of the electron distribution function is large enough, and formulae \eqref{eq:nu_tau} and \eqref{eq:tau} are no longer applicable. Indeed, in moderate temperatures, the viscosity is described by the more general Eq.~\eqref{eq:nu_general}. The temperature smearing of the Fermi surface is accounted in the numerical computation of the viscosity in Ref.~\cite{keser2021}. The result of this calculation shows that, already at $k_\mathrm{B}T/E_\mathrm{F} \approx 0.1$, the viscosity temperature dependence deviates significantly from $1/T^{2}$, and that is observed in our experiment.

Note also that, in the whole temperature range, the $\nu$ values in suspended samples are lower than in non-suspended ones. In particular, at $k_\mathrm{B}T \ll E_\mathrm{F}$ (corresponds to $T \lesssim 10$~K in our case), the formula \eqref{eq:nu_tau} is applicable and
$$\frac{\nu_\mathrm{non\text{-}susp}}{\nu_\mathrm{susp}} = \frac{\tau_\mathrm{ee,\,non\text{-}susp}}{\tau_\mathrm{ee,\,susp}}=\frac{W_\mathrm{ee,\,susp}}{W_\mathrm{ee,\,non\text{-}susp}}\approx 1.4,$$
where $W_\mathrm{ee}=1/\tau_\mathrm{ee}$ is the e-e scattering probability. The increased scattering probability is explained by the e-e interaction enhancement in suspended structures due to the removal of part of the polarizing medium from PC \cite{pogosov2022}, and that changes an effective dielectric constant value $\varepsilon$. The viscosities ratio before and after the suspension monotonically grows from 1.4 to 1.6 with the temperature increase from 4 to 90~K (inset in Fig.~\ref{fig:viscosity}). The viscosities ratio growth may be due to the non-trivial dependence of $\tau_\mathrm{ee}$ on $\varepsilon$ \cite{giuliani1982}. So, the sample suspension, on the one hand, does not affect the scattering on phonons and impurities (pay attention to the identical $R_\mathrm{2DEG}(T)$ dependence in Fig.~\ref{fig:resistance}~(c) and Fig.~\ref{fig:resistance}~(d)) and, on the other hand, it reduces $l_\mathrm{ee}=v_\mathrm{F}\tau_\mathrm{ee}$; it means that the suspension reduces the lower temperature bound for the hydrodynamic effects observation. As a consequence, suspended systems are preferred for the study of electron hydrodynamics.

\section{Conclusion}
We experimentally demonstrated,
that electron transport in both suspended and non-suspended GaAs PCs of different width can be described in the framework of the viscous flow with unusual superballistic conductance that increases with temperature. The viscous contribution to the PC conductance, extracted from the experiment, is proportional to the square of the PC width ($G_\mathrm{vis}\propto w^2$) that, according to the theory \cite{guo2017}, confirms its viscous origin. The electron viscosity $\nu$, that is inversely proportional to the viscous contribution, has the dependence $1/T^2$ at low temperatures ($k_\mathrm{B}T\ll E_\mathrm{F}$) and follows $1/T$ at moderate temperatures. The low-$T$ dependence is in agreement with the theory predicting that $\nu\propto\tau_\mathrm{ee}\propto1/T^2$, while the divergence with the theory at moderate temperatures can be due to the temperature smearing of the electron distribution function. The viscosity temperature dependence is shown to have the same character after the suspension, but the viscosity value is reduced due the to e-e interaction enhancement in suspended structures of about 1.4~times.

\begin{acknowledgments}
The study was funded by the Russian Science Foundation (grant No. 22-12-00343).
\end{acknowledgments}

\bibliography{mybib}

\end{document}